\documentclass[a4paper, 9pt, twocolumn]{article}
\usepackage{geometry}                
\usepackage{graphicx}
\usepackage{amssymb}
\usepackage{amsthm}
\usepackage{epstopdf}
\usepackage{mathtools}
\usepackage[shortlabels]{enumitem}
\usepackage{dsfont}
\usepackage{cuted}
\usepackage{sidecap}
\usepackage[usenames,dvipsnames]{color}
\usepackage[thinspace,mediumqspace,Grey,squaren]{SIunits}
\usepackage{bbm}
\usepackage[T1]{fontenc}
\usepackage[utf8]{inputenc}
\usepackage{authblk}
\usepackage[font=small,labelfont=bf ,textfont={small, sf}]{caption}
\usepackage{titlesec}
\usepackage{url}
\usepackage{hyperref}
\usepackage{algorithm}
\usepackage{algorithmic}
\titleformat{\subsection}[runin]{\normalfont\bfseries}{\thesubsection.}{3pt}{}

\begin{document}
\title{Sparse Learning of Markovian Population Models in Random Environments}

\author[1]{Christoph Zechner}
\author[1]{Federico Wadehn}
\author[1,2,3]{Heinz Koeppl}

\affil[1]{Automatic Control Lab, ETH Zurich, 8092 Zurich, Switzerland}
\affil[2]{IBM Zurich Research Laboratory, 8803 R\"uschlikon, Switzerland}
\affil[3]{Correspondence to: koeppl@ethz.ch}
\date{}
\maketitle
\newpage{}

{\noindent \sf \fontsize{9}{0} \selectfont \textbf{
Markovian population models are suitable abstractions to describe well-mixed interacting particle systems in situation where stochastic fluctuations are significant due to the involvement of low copy particles. In molecular biology, measurements on the single-cell level attest to this stochasticity and one is tempted to interpret such measurements across an isogenic cell population as different sample paths of one and the same Markov model. Over recent years evidence built up against this interpretation due to the presence of cell-to-cell variability stemming from factors other than intrinsic fluctuations. To account for this extrinsic variability, Markovian models in random environments need to be considered and a key emerging question is how to perform inference for such models. We model extrinsic variability by a random parametrization of all propensity functions. To detect which of those propensities have significant variability, we lay out a sparse learning procedure captured by a hierarchical Bayesian model whose evidence function is iteratively maximized using a variational Bayesian expectation-maximization algorithm.    
}}



\newcommand{\vect}[1]{\mathbf{#1}}
\newcommand{\vectSymbol}[1]{\boldsymbol{#1}}

\newcommand{\Normal}[2]{\mathcal{N}\left( #1, #2 \right)}

\newcommand{\LowerBound}[1]{\mathcal{L}\left[ #1 \right]}
\newcommand{\KL}[2]{KL\left[ #1 \| #2 \right]}


\def\MCChain{\Lambda}
\def\mCChain{\lambda}
\def\mCChainInt{\boldsymbol{\lambda}}
\def\MCIndex{j}

\def\Time{t}
\def\TimeIndex{N}
\def\timeIndex{l}
\def\DiscreteTime{\Time_{\timeIndex}}

\def\InitialDistribution{\pi}
\def\ParameterPrior{\phi}
\def\SingleParamPrior{\phi}
\def\MeasurementParamPrior{\psi}
\def\SingleMeasurementParamPrior{\psi}

\def\MeasurementSigma{\Sigma}
\def\measurementSigma{\sigma}

\def\Joint{p}
\def\Density{p}
\def\Proposal{q}

\def\GammaDistribution{\mathcal{G}}

\def\noise{u}
\def\Noise{U}

\def\NExtrinsic{N}
\def\extrinsicVariable{z}
\def\ExtrinsicVariable{Z}
\def\ExtrinsicSpace{\mathcal{Z}}
\def\hyperParameters{a}
\def\HyperParameters{A}
\def\HyperSpace{\mathcal{A}}
\def\hyperHyperParameters{b}
\def\HyperHyperParameters{B}

\def\MarkovChain{\State}
\def\State{X}
\def\StatePopulation{\mathbb{X}}
\def\StateInt{\mathbf{\State}}
\def\state{x}
\def\stateInt{\mathbf{\state}}
\def\StateTime{\State({\Time})}
\def\stateDiscreteTime{\state_{\timeIndex}}
\def\StateDim{d}
\def\ReactionDim{\nu}
\def\CellCount{M}
\def\StateSpace{\mathbb{N}_{+}}
\def\StateSpaceDim{\StateSpace^{\StateDim}}

\def\Measurement{Y}
\def\MeasurementInt{{\Measurement}}
\def\measurement{y}
\def\measurementInt{{\measurement}}
\def\MeasurementTime{\Measurement_{\timeIndex}}
\def\measurementTime{\measurement_{\timeIndex}}
\def\MeasurementDim{o}
\def\MeasurementSpace{\mathbb{R}}
\def\MeasurementSpaceDim{\MeasurementSpace^{\MeasurementDim}}

\def\PathDensity{f}
\def\MeasurementDensity{g}

\def\AllParameters{\Theta}
\def\allParameters{\theta}

\def\SharedParameters{S}
\def\sharedParameters{s}

\def\kParameters{C}
\def\kparameters{c}
\def\kparameter{c}

\def\Lapvar{\sigma}

\def\MargProp{h}

\def\Parameters{C}
\def\parameters{c}
\def\parameter{c}

\def\MeasurementParameters{\Omega}
\def\measurementParameters{\omega}

\def\Covariates{V}
\def\covariates{v}

\def\GammaDist{\mathcal{G}}

\def\df{\mathrm{d}}

\def\MarkovBlanket{\mathcal{MB}}

\def\History{\mathcal{H}}

\def\Real{\mathbb{R}}
\def\RealPos{\mathbb{R}_{\geq 0}}

\def\FullLatentSet{U}
\def\fullLatentSet{u}
\def\FullMeasurementSet{Z}
\def\fullMeasurementSet{z}
\def\LatentSubSet{\FullLatentSet}
\def\latentSubSet{\fullLatentSet}

\newcommand{\ExpectSub}[2]{ \mathbb{ E }_{#1} \left[ #2 \right]}
\newcommand{\Expect}[1]{ \mathbb{ E } \left[ #1 \right]}
\newcommand{\ExpectWRT}[2]{ \mathbb{ E }_{ #2 } \left[ #1 \right]}
\newcommand{\Var}[1]{ \mathrm{Var} \left[ #1 \right]}
\newcommand{\VarWRT}[2]{ \mathrm{ Var }_{ #2 } \left[ #1 \right]}

\def\Mgf{G}

\def\PriorParameters{\Theta}
\def\priorParameters{\theta}
\def\priorParamA{\kappa}
\def\priorParamB{\chi}
\def\PathSpace{\mathcal{X}}

\newcommand{\LogNormal}[2]{\mathcal{LN} \left( #1, #2 \right)}
\newcommand{\Uniform}[2]{\mathcal{U} \left( #1, #2 \right)}

\def\Model{\mathcal{M}}
\def\model{m}
\def\BayesFactor{\mathcal{K}}

\def\IndexSet{\mathcal{I}}
\def\CylinderSet{\Gamma(\stateInt)}

\def\MorphParameters{B}
\def\morphParameters{b}

\newcommand{\Indicator}[2]{\mathbbm{1}_{#2} \left( #1 \right)}

\newcommand{\Comment}[2]{ {\color{blue} #1} \textit{Remark: #2}}

\section{Introduction}
Markovian population models are ubiquitous in biology to capture the temporal change in abundance for different particle types (i.e. species) caused by interactions or transformations among them. Inferring such models from experimental data is at the core of quantitative biology. Reconstructing models of biochemical cellular processes using the principles of chemical kinetics is an important example. Single-cell technologies provide unprecedented means to perform this task, however novel computational methods are required to deal with the complexity of single-cell data. More specifically, such data represent a heterogeneous aggregate of measurements due to the fact that cells are not exactly identical to start with. Thus, apart from stochastic fluctuations intrinsic to the process under study, {\it extrinsic} sources of variability contribute to the overall heterogeneity (\cite{Elowitz2002, Colman-Lerner2005}). The single-cell process gets modulated by its local microenvironment, that can refer to intracellular quantities such as initial copy numbers of participating biomolecules (\cite{Koeppl2012}) but also cell-level quantities like the cell's local growth condition (\cite{Snijder2011}), or its cell-cycle stage (\cite{Colman-Lerner2005}). To capture this variability in a computational model is challenging because the true sources and their strength for a specific cell line are yet to be identified. Hence, recent approaches to address extrinsic noise in the inference procedure have to make an educated guess which quantities of a kinetic model are modulated by extrinsic variability (\cite{Zechner2012,Zechner2014, Ruess2013, Hasenauer2011,Shahrezaei2008}). For instance, one source of extrinsic noise in gene expression that is believed to be significant are ribosome copy number variations. Taking this as a starting point, we recently developed an inference framework that relies on a hierarchical Bayesian model, where some latent states express the extrinsic variability of an actual model quantity (\cite{Zechner2014}). The structure of this Bayesian model is fixed beforehand and hence can not be changed {\it a posteriori} when the data is incorporated. 

Here we lay out an inference framework where the hierarchical dependency structure among model quantities and extrinsic sources is learned from the data. In order to retrieve results that are interpretable and robust with respect to small sample sizes (i.e. number of cells) we apply a sparse Bayesian learning technique (\cite{Neal1996, Tipping2001, Bishop2007}) yielding the named dependency structure with a minimal number of edges. In order to reduce the number of unknown parameters, we make use of the marginalized process introduced in \cite{Zechner2014}. To infer the posterior with respect to the extrinsic variability in an efficient manner we employ a variational Bayesian expectation-maximization (EM) procedure (\cite{Beal2003, Shutina, Dempster1977}). The outlined method assumes the availability of data in terms of complete and noise free sample paths. The method can be generalized to the more realistic incomplete and noisy data case but the necessary computational machinery for that would sidetrack the exposition and occlude the main idea behind this approach. To this end, the work represents a first step towards a model-based understanding of how and which concurrent processes modulate a specific cellular process under study {\it in vivo}.        

The remaining part of the paper is structured as follows. In Section \ref{sec:Math} we derive the mathematical models and algorithms. We start with a brief introduction to heterogeneous population models (Section \ref{sec:StochChemKin}). Subsequently, in Section \ref{sec:HierarchicalBayesianModeling} we develop a suitable hierarchical Bayesian model whose inference is discussed in Section \ref{sec:VariationalInference}. In Sections \ref{sec:Implementation} and \ref{sec:IncompleteData} we address several practical aspects of the algorithm and briefly discuss how it extends to the incomplete data scenario. The algorithm is analyzed and validated in Section \ref{sec:Simulations} using a few case studies.

\section{Mathematical Framework}
\label{sec:Math}

\subsection{Notation}
Random quantities and their realizations are denoted by upper- and lowercase symbols, respectively. Symbol $p$ and $q$ are used to indicate the exact and the approximating probability density functions (PDFs), respectively and expectations are denoted by $\Expect{A} = \int a p(a) \mathrm{d}a$. For convenience we also introduce expectations of the form $\ExpectSub{a}{f(a, b)} = \int f(a, b) p(a) \mathrm{d}a$, indicating that the expectation of $f$ is only taken with respect to $p(a)$. We denote the Gamma distribution by $\GammaDistribution(\alpha, \beta)$ with $\alpha$ and $\beta$ as shape and inverse scale parameters. The exponential distribution is denoted $\mathrm{Exp}(\lambda)$ with inverse scale parameter $\lambda$. Furthermore, we express time-dependent quantities on intervals $[0, T]$ by bold symbols, e.g., $\stateInt = \{ x(t) \mid 0 \leq t \leq T \}$. The symbol $KL[q(x) \| p(x)]$ denotes the Kullback-Leibler divergence between PDFs $q(x)$ and $p(x)$, i.e., $KL[q(x) \| p(x)] = \int q(x) \ln\frac{q(x)}{p(x)} \mathrm{d}x$. Abbreviations CV and SCV stand for coefficient of variation and squared coefficient of variation, respectively.

\subsection{Population Models and Chemical Kinetics}
\label{sec:StochChemKin}

We consider a continuous-time Markov chain (CTMC) $\MarkovChain$ describing the dynamics of a stochastic interaction network comprising $\StateDim$ species and $\ReactionDim$ coupled reaction or transformation channels. The latter are associated with a set of real-valued kinetic parameters $\kParameters = \{ \kParameters_{j} \mid j=1,\ldots, \ReactionDim\}$. The species' abundances at time $t$ defines the random state $\MarkovChain(t) = x$, $x \in \mathbb{Z}_{\geq 0}^\StateDim$ of the network. Propensity functions corresponding to each channel are general functions of the state and can often be defined through first principles such as the law of mass-action. Throughout the work we assume propensities to be linear functions of their respective rate parameters, i.e. they take the general form $c_i g_i(x)$, where $g_i(x)$ is an arbitrary nonlinear function.  Under knowledge of the parameters $\kParameters$, the dynamics of a single cell follows a conditional CTMC $\MarkovChain \mid \kParameters$.

Due to extrinsic cell-to-cell variability, acquired single-cell trajectories can not be thought of as being different realizations of a single CTMC. Although sources of extrinsic variability can be diverse, we throughout the work make the assumption that kinetic parameters are the only source of extrinsic variability that enters the cellular process under study. In contrast to previous approaches where specific parameters were subject to extrinsic variability (\cite{Zechner2014, Zechner2012}), we assign prior variability to every kinetic parameter in the model.  
%
Accordingly, we associate to $\kParameters$ a probability distribution, i.e., $\kParameters \mid (\HyperParameters=\hyperParameters)  \sim p(\kparameters \mid \hyperParameters)$, with $\HyperParameters$ a set of hyperparameters. With this, the dynamics of the $m$-th cell of a population is described by a conditional CTMC $\MarkovChain^{m} \mid (\kParameters^{m}  = \kparameters)$. We remark that the parameter dimensionality of the heterogeneous CTMC model increases with every considered cell and hence, scales poorly with the population size $M$. Fortunately, it was recently shown that a CTMC $\MarkovChain \mid \kParameters$ can be integrated over $\kParameters$, yielding a marginalized stochastic process $\MarkovChain \mid \HyperParameters$, which directly depends on the hyperparameters $\HyperParameters$. While fuller details about the construction and simulation of such a process can be found in \cite{Zechner2013b,Zechner2014,AalenBook}, we only introduce the key quantity needed here, i.e., the marginal path likelihood function. We know from \cite{Wilkinson2006} and \cite{Kuechler1997} that the path likelihood function of an observed sample path $\stateInt = \{x(t) \mid 0\leq t\leq T  \}$ is given by
\begin{equation}
	p(\stateInt \mid \kparameters) \propto \prod_{i=1}^{\ReactionDim} \kparameter_{i}^{r_{i}(\stateInt)} e^{-\kparameter_{i} \int_{0}^{T} g_{i}(x(t)) \mathrm{d}t},
\end{equation}
with $r_{i}(\stateInt)$ the number of reactions of type $i$ that happened in the path $\stateInt$. Formally, the marginal path likelihood is obtained via the integral
\begin{equation}
	p(\stateInt \mid \hyperParameters) = \int p(\stateInt \mid \kparameters) p(\kparameters \mid \hyperParameters) \mathrm{d}\kparameters,
\end{equation}
whose tractability depends on $p(\kparameters \mid \hyperParameters)$. For instance, the Gamma distribution was shown to have convenient analytical properties (\cite{Zechner2014}) and furthermore, appears plausible in the context of gene expression \cite{Xie2010}. Throughout the remaining paper we assume
\begin{equation}
	p(\kparameters \mid \hyperParameters) = \prod_{i=1}^{\ReactionDim} \GammaDistribution(\kparameter_{i} \mid \alpha_{i}, \beta_{i}),
	\label{eq:GammaAssumption}
\end{equation}
with $\hyperParameters = \{ (\alpha_{i}, \beta_{i} ) \mid i=1,\ldots,\ReactionDim \}$ and $\GammaDistribution(\kparameter_{i} \mid \alpha_{i}, \beta_{i})$ as a Gamma distribution over $\kparameter_{i}$ with shape- and inverse scale parameters $\alpha_{i}$ and $\beta_{i}$, respectively. A suitable measure of a reaction channel's extrinsic variability is the squared coefficient of variation (or normalized variance), which in the Gamma-case is given by
\begin{equation}
	\eta_{i} = \frac{1}{\alpha_{i}}, \nonumber
\end{equation}
indicating that one can detected heterogeneity by merely analyzing the shape parameter $\alpha_{i}$. Under assumption (\ref{eq:GammaAssumption}), the marginal path likelihood function is given by (see e.g., \cite{Zechner2014})
\begin{equation}
	\begin{split}
	 &p(\stateInt \mid \hyperParameters) = \prod_{i=1}^{\ReactionDim} p(\stateInt \mid \alpha_{i}, \beta_{i}) \\
	 &\propto \prod_{i=1}^{\ReactionDim} \frac{\beta_{i}^{\alpha_{i}}\Gamma(\alpha_{i} + r_{i}(\stateInt))}{\Gamma(\alpha_{i})} 
	\left( \beta_{i} + \int_{0}^{T} g_{i}(x(t)) \mathrm{d}t \right)^{-(\alpha_{i} + r_{i}(\stateInt))}.
	\end{split}
	\label{eq:MarginalLikelihood}
\end{equation}
Due to the marginalization, the hidden layer corresponding to the kinetic parameters is entirely removed and hence, does not need to be considered in the following derivations.

\subsection{Hierarchical Bayesian Modeling}
\label{sec:HierarchicalBayesianModeling}

\begin{figure}[ht!]
\begin{center}
	\includegraphics[width=0.9\columnwidth]{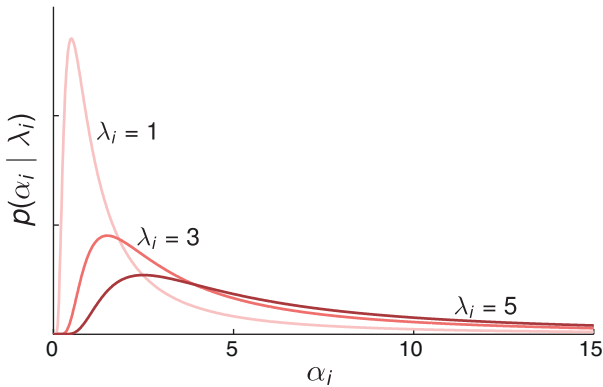}
\caption{Prior distributions over $\alpha_{i}$ for different values of the hyperparameter $\lambda_{i}$. The distributions show a peak for low values of $\alpha_{i}$ and become more heavy-tailed with increasing $\lambda_{i}$.}
\label{fig:PriorDistributions}
\end{center}
\end{figure}

Assume we have given measurements of $M$ cells of a heterogeneous population, i.e., $\stateInt^{m}$ for $m=1,\ldots,M$. The extrinsic variability of each reaction channel $i$ can be quantified by inferring the hyperparameters $\{\alpha_{i}, \beta_{i}\}$ from those measurements. According to a Bayesian scenario, this is equivalent to finding the posterior distribution
\begin{equation}
	\begin{split}
	p(\hyperParameters \mid \stateInt^{1}, \ldots, \stateInt^{M}) &\propto \prod_{m=1}^{M} p(\stateInt^{m} \mid \hyperParameters) p(\hyperParameters) \\
		&=\prod_{m=1}^{M} \left(  \prod_{i=1}^{\ReactionDim} p(\stateInt^{m} \mid \alpha_{i}, \beta_{i}) p(\alpha_{i}, \beta_{i}) \right).
	\end{split}
\end{equation}
which hence, factorizes such that
\begin{equation}
	p(\hyperParameters \mid \stateInt^{1}, \ldots, \stateInt^{M}) = \prod_{i=1}^{\ReactionDim} p(\alpha_{i}, \beta_{i} \mid \stateInt^{1}, \ldots, \stateInt^{M}).
	\label{eq:PosteriorFact}
\end{equation}
Naively, one could just evaluate the individual terms in (\ref{eq:PosteriorFact}) and check whether the corresponding values of $\alpha_{i}$ are below a certain threshold, indicating heterogeneity of the associated reaction. However, since those values are only accessible through the noisy measurements $\stateInt^{m}$, it is not clear how to choose such a threshold in order to obtain maximally robust results. For instance, the heterogeneity stemming from the intrinsic molecular fluctuations should be "filtered out" and yield a negative detection result. Positive detections are only desired if there is significant evidence in the data. Technically, this corresponds to solving a $\textit{sparse}$ Bayesian learning problem (\cite{Neal1996, Tipping2001, Bishop2007}). 
The key step to achieve sparsity in empirical Bayesian models is to assign suitable prior - and hyperprior distributions to the model quantities. Since detection of heterogeneity is based on only $\alpha_{i}$, we chose 
\begin{equation}
	p(\alpha_{i}, \beta_{i}) = p(\alpha_{i} \mid \lambda_{i}) p(\beta_{i}),
\end{equation}
where $\lambda_{i}$ controls the shape of $p(\alpha_{i} \mid \lambda_{i})$ and $p(\beta_{i})$ is assumed to be flat over the positive domain, such that $p(\alpha_{i}, \beta_{i}) \propto p(\alpha_{i} \mid \lambda_{i})$.
The goal is to define $p(\alpha_{i} \mid \lambda_{i})$ such that the heterogeneity is forced to zero unless there is significant evidence in the data. Accordingly, suitable distributions will emphasize SCVs around zero while also permitting high values. Here we choose $p(\alpha_{i} \mid \lambda_{i})$ such that $p(\eta_{i}) = \mathrm{Exp}(\lambda_{i})$. A transformation of random variables yields
\begin{equation}
	p(\alpha_{i} \mid \lambda_{i}) =  \frac{\lambda_{i}}{\alpha_{i}^{2}} e^{-\frac{\lambda_{i}}{\alpha_{i}}}.
\end{equation}
The resulting prior distributions over $\alpha_{i}$ are illustrated in Fig.~\ref{fig:PriorDistributions} for different values of $\lambda_{i}$.

While standard Bayesian approaches rely on \textit{given} prior knowledge, empirical Bayes techniques aim to infer parameters as well as their hyperparameters from data. In our case, this means that in addition to $\alpha_{i}$ and $\beta_{i}$, also the hyperparameters $\lambda_{i}$ are assumed to be unknown and need to be estimated. In order to obtain a fully Bayesian model, we need to specify hyperprior distributions $p(\lambda_{i})$. Again, we assume $p(\lambda_{i})$ to be flat but remark that an extension to arbitrary distributions is straight-forward.
With the model parameters $\hyperParameters$ and their hyperparameters $\hyperHyperParameters = \{ \lambda_{i} \mid i=1,\ldots,\ReactionDim \}$, we aim to compute the posterior distribution
\begin{equation}
	\begin{split}
		&p(\hyperParameters, \hyperHyperParameters \mid \stateInt^{1}, \ldots, \stateInt^{M}) \\
		&\quad \quad \propto \prod_{i=1}^{\ReactionDim} \left( \prod_{m=1}^{M}  p(\stateInt^{m} \mid \alpha_{i}, \beta_{i}) \right)  p(\alpha_{i} \mid \lambda_{i}) p(\lambda_{i}) \\
		&\quad \quad = \prod_{i=1}^{\ReactionDim} p(\alpha_{i}, \beta_{i}, \lambda_{i} \mid \stateInt^{1}, \ldots, \stateInt^{M})
	\end{split}
	\label{eq:TargetPosterior}
\end{equation}
where the r.h.s. of (\ref{eq:TargetPosterior}) is just the joint distribution over all model quantities. Unfortunately, it turns out that (\ref{eq:TargetPosterior}) is intractable. In the next section we will develop an variational inference scheme to approximate (\ref{eq:TargetPosterior}). 

\subsection{Variational Inference}
\label{sec:VariationalInference}

Variational inference schemes aim to approximate some target posterior $p(z\mid y)$ by some other distribution $q(z)$. More specifically, one chooses $q(z)$ such as to minimize the Kullback-Leibler divergence (KL) between $q(z)$ and the true distribution. For that sake, note that for every $q$, the log-evidence function satisfies the decomposition (\cite{Beal2003, Bishop2007})
\begin{equation}
	\ln p(z) = \LowerBound{q(z)} + \KL{q(z)}{p(z\mid y)},
	\label{eq:LogEvidence}
\end{equation}
where $\LowerBound{q(z)}$ forms a lower bound on $\ln p(z)$ which is given by
\begin{equation}
\LowerBound{q(z)} = \int q(z) \ln \frac{p(z, y)}{q(z)} \mathrm{d}z.
\label{eq:LowerBound}
\end{equation}
Accordingly, minimizing the KL with respect to $q$ is the same as maximizing its counterpart $\LowerBound{q(z)}$, i.e., 
\begin{equation}
	q^{*}(z) = \underset{q(z)\in Q}{\mathrm{argmax}}~\LowerBound{q(z)}.
\end{equation}
It can be seen from (\ref{eq:LogEvidence}) and (\ref{eq:LowerBound}) that $\LowerBound{q(z)}$ is maximal if and only if $q(z)=p(z\mid y)$. In order to obtain a tractable $q(z)$, one typically imposes further constraints on its structure. Most commonly, individual components of $z$ are assumed to be independent of each other, i.e., 
\begin{equation}
	q(z) = \prod_{l=1}^{L} q(z_{l}),
\end{equation}
also known as the \textit{mean-field} approximation (\cite{Beal2003}). In this case, it can be shown that the optimal variational solution of the individual factors $q(z_{i})$ is determined by
\begin{equation}
	\ln q^{*}(z_{i}) = \ExpectSub{j\neq i}{\ln p(z, y)} + const.
	\label{eq:FreeFormSolution}
\end{equation}
where $\ExpectSub{j\neq i}{\ln p(z, y)}$ denotes the expectation of the logarithm of the joint distribution, taken with respect to all factors $q(z_{j})$ except $q(z_{i})$. Since the optimal solution of a particular $q$-factor depends on all other factors, the mean-field approximation typically induces an iterative inference scheme, where the individual factors are updated in a round-robin fashion. Such schemes stand in close relation with traditional expectation-maximization (EM) algorithms (\cite{Dempster1977}) and accordingly, are often referred to as variational Bayesian EM (VBEM) algorithms (\cite{Shutina, Beal2003}).

In practice, eq. (\ref{eq:FreeFormSolution}) might still be intractable, in which case it is necessary to further restrict the corresponding $q$-factor. For instance, one could assume $q(z_{i})$ to be some parameterized distribution (e.g., a Gaussian with mean and variance) and determine its parameters $\theta$ as
\begin{equation}
	\theta^{*} = \underset{\theta \in \Theta}{\mathrm{argmax}}~\Expect{\ln p(z, y)},
\end{equation}
whereas in this case, the expectation is taken with respect to all $q$-factors. For instance, if one is interested solely in maximum a-posterior (MAP) estimates, $q(z_{i})$ can be chosen to be a Dirac-delta function with unknown position. 

We will now use the VBEM framework to derive an approximate iterative inference algorithm for the hierarchical Bayesian model from Section \ref{sec:HierarchicalBayesianModeling}. The goal is to compute an approximate posterior distribution $q(\hyperParameters, \hyperHyperParameters)$ for which we assume that it factorizes as
\begin{equation}
	q(\hyperParameters, \hyperHyperParameters) = \prod_{i=1}^{\ReactionDim} q(\alpha_{i}, \beta_{i}) q(\lambda_{i}).
\end{equation}
We remark that the in the complete-data scenario considered here, also the true posterior factors over the individual reaction channels $i=1,\ldots,\ReactionDim$, however, not over $\{\alpha_{i}, \beta_{i} \}$ and $\lambda_{i}$. For analytical simplicity, we further assume $q(\lambda_{i}) := \delta(\lambda_{i} - \hat{\lambda}_{i})$ with $\hat{\lambda}_{i}$ as an unknown position parameter.
The factor $q(\alpha_{i}, \beta_{i})$ for the $i$-th reaction channel is determined by
\begin{equation}
	\begin{split}
		& \ln q^{*}(\alpha_{i}, \beta_{i}) = \ExpectSub{\lambda_{i}}{\ln p(\hyperParameters, \hyperHyperParameters, \stateInt^1, \ldots, \stateInt^M)} + const.,
		\end{split}
		\end{equation}
which becomes 
\begin{equation}
	\begin{split}
			\ln q^{*}(\alpha_{i}, \beta_{i})  &= \sum_{m=1}^{M}  \ln p(\stateInt^{m} \mid \alpha_{i}, \beta_{i}) \\
			&\quad+ \ExpectSub{\lambda_{i}}{\ln p(\alpha_{i} \mid \lambda_{i})} + const.
	\end{split}
\end{equation}
when taking into account the r.h.s. of eq. (\ref{eq:TargetPosterior}). Together with the marginal path-likelihood function from eq.~(\ref{eq:MarginalLikelihood}), we further obtain
\begin{equation}
\begin{split}
	&\ln q^{*}(\alpha_{i}, \beta_{i})  \\
	&\quad \quad = \sum_{m=1}^{M}\alpha_{i} \ln \beta_{i} + \ln \Gamma(\alpha_{i} + r_{i}(\stateInt^{m})) - \ln\Gamma(\alpha_{i}) \\
	&\quad \quad \quad -(\alpha_{i} + r_{i}(\stateInt^{m}))  \ln \left( \beta_{i} + \int_{0}^{T} g_{i}(x^{m}(t)) \mathrm{d}t \right) \\
	&\quad\quad\quad -\frac{ \hat{\lambda}_{i}}{\alpha_{i}} - 2 \ln \alpha_{i} + const.,
	\label{eq:LogPosteriorAlphaBeta}
\end{split}
\end{equation}
where we have used the fact that $$\ExpectSub{\lambda_{i}}{\lambda_{i}} = \int \lambda_{i} \delta(\lambda_{i} - \hat{\lambda}_{i}) \mathrm{d}\lambda_{i} = \hat{\lambda}_{i}.$$ Although eq. (\ref{eq:LogPosteriorAlphaBeta}) is not of standard form, it can be evaluated analytically or using a suitable sampling algorithm. 

The $q$-factor corresponding to $\lambda_{i}$ is found by solving the parametric (instead of variational) optimization
\begin{equation}
	\begin{split}
		\lambda_{i}^{*} &= \underset{\hat{\lambda}_{i} \in \Real}{\mathrm{argmax}}~\Expect{\ln p(\hyperParameters, \hyperHyperParameters, \stateInt^1, \ldots, \stateInt^M)} \\
		&=\underset{\hat{\lambda}_{i} \in \Real}{\mathrm{argmax}}~\Expect{\ln p(\alpha_{i} \mid \lambda_{i} )}.
	\end{split}
	\label{eq:LambdaOptimization}
\end{equation}
The expectation inside the maximum operator is given by
\begin{equation}
	\Expect{\ln p(\alpha_{i} \mid \lambda_{i} )} = - \hat{\lambda}_{i} \ExpectSub{\alpha_{i}}{\alpha_{i}^{-1}} + \ln \hat{\lambda}_{i} - 2 \ExpectSub{\alpha_{i}}{\ln \alpha_{i}},
\end{equation}
whose maximum is found to be
\begin{equation}
	\lambda_{i}^{*}= \frac{1}{\ExpectSub{\alpha_{i}}{\alpha_{i}^{-1}}}.
	\label{eq:LambdaOpt}
\end{equation}

\subsection{Implementation Aspects}
\label{sec:Implementation}
As mentioned earlier, the VBEM scheme leads to an iterative algorithm, where each $q$-factors are estimated successively, given the most recent estimates of all other q-factors. For a particular reaction channel $i$, this means that we first determine $q(\alpha_{i}, \beta_{i})$ given the most recent value of $\hat{\lambda}_{i}$ and subsequently re-estimate $\hat{\lambda}_{i}$ given $q(\alpha_{i}, \beta_{i})$ and so forth. Since $q(\alpha_{i}, \beta_{i})$ is not of standard form, we can compute its required statistics either via numerical integration or Monte Carlo sampling. Here we focus on the latter approach and employ a standard Metropolis-Hastings (M-H) sampler with log-normal proposal distributions to draw samples from $q(\alpha_{i}, \beta_{i})$. Those samples are also used for updating the corresponding hyperparameters $\lambda_{i}$, i.e., the expectation in eq. (\ref{eq:LambdaOpt}) is replaced by a Monte Carlo average. Moreover, we found that replacing $\ExpectSub{\alpha_i}{\alpha_i^{-1}}$ by $\ExpectSub{\alpha_i}{\alpha_i}^{-1}$ yields a similar estimation performance, while significantly reducing the number of required divisions per iteration.

Note that the parameters corresponding to the homogeneous reaction channels will be driven to infinity, which in theory, causes the algorithm to diverge. Practically -- however -- one can check whether $\alpha_i$ (or $\lambda_i$) is above a critical threshold (e.g., around $10e5$), in which case the $i$-th reaction is considered homogeneous and excluded from the remaining analysis.

Algorithm \ref{alg:VBEM} summarizes the main structure of the proposed scheme.

\begin{algorithm}
\caption{VBEM algorithm for detecting heterogeneity in stochastic interaction networks.}
\label{alg:VBEM}
	\begin{algorithmic}[1]
		\STATE Initialize $\hat{\lambda}_{i}$ for $i=1, \ldots, \ReactionDim$
		\WHILE{not converged} 
			\FOR{$i=1,\ldots,\ReactionDim$} 
				\STATE Draw samples from $q(\alpha_{i}, \beta_{i})$ using eq. (\ref{eq:LogPosteriorAlphaBeta}) and the current value of $\hat{\lambda}_{i}$
				\STATE Update $\hat{\lambda}_{i}$ using eq. (\ref{eq:LambdaOpt})
			\ENDFOR 
		\ENDWHILE
	\end{algorithmic}		
\end{algorithm}

\subsection{Extension to the Incomplete Data Scenario}
\label{sec:IncompleteData}
In principle, the above algorithm can be easily extended for the incomplete data scenario, i.e., if the measurements consist of sparse and noisy readouts $\Measurement_{n}$ of the Markov chain $\MarkovChain$ at times $t_{n}$. Intuitively, this can be understood as adding another layer on top of the states $\stateInt^{m}$ in the hierarchical Bayesian model. In this case it turns out that the variational expressions from Section \ref{sec:VariationalInference} also involve expectations with respect to so-called \textit{smoothing} distributions, e.g., $p(\stateInt^{m} \mid \measurement_{1}^m, \ldots, \measurement_{N}^m, \hyperParameters)$ when considering the $m$-th cell. Computing such distributions (and computing its statistics) is a challenging task on its own and a variety of numerical and analytical approaches have been proposed (\cite{Amrein2012, Opper2007, Zechner2014}). Apart from that, the VBEM framework can be readily applied to the more complicated case of incomplete and noisy measurements.

\section{Simulations}
\label{sec:Simulations}
\begin{figure}[ht!]
\begin{center}
	\includegraphics[width=0.99\columnwidth]{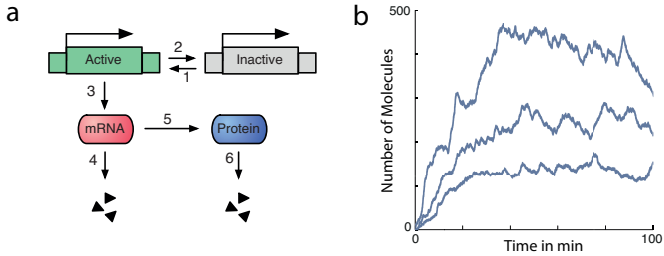}
\caption{A simple model of eukaryotic gene expression. \textbf{(a)} Schematic diagram of the reaction network. The numbered arrows indicate chemical events taking place: upon activation of the gene (arrow 1), mRNA can be transcribed (arrow 3) which in turn gets translated (arrow 5) into protein. The remaining arrows indicate gene-deactivation (arrow 2) and degradation events (arrows 4 and 6). \textbf{(b)} Exemplary protein traces of a heterogeneous network. In this case, heterogeneity was simulated by introducing a Gamma-type variability in the translation rate.}
\label{fig:Model}
\end{center}
\end{figure}

We performed several simulation studies in order to demonstrate and evaluate the proposed method. For each of the case studies, we used the simple reaction network of eukaryotic gene expression illustrated in Fig.~\ref{fig:Model}a. Exemplary trajectories of such a model are shown in Fig.~\ref{fig:Model}b. The model comprises six reaction channels with kinetic parameters $\kparameters_{1}, \ldots, \kparameters_{6}$, which are either homogenous or heterogeneous -- depending on the particular case study. 

Unless otherwise specified, the mean values of the kinetic parameters are chosen according to Table~\ref{tab:ParamTable}.

\begin{table}[ht!]
\caption{Mean values of the kinetic parameters.}
\begin{center}
\begin{tabular}{c|c c c c c c}
Parameter 			&$\kparameter_1$ &$\kparameter_2$ &$\kparameter_3$ 	&$\kparameter_4$    &$\kparameter_5$ &$\kparameter_6$\\ \hline
Mean ($s^{-1}$)         &$0.5$ 		     &$0.05$	 	  &$0.1$			&$0.001$			&$0.03$		           &$0.008$\\ 
\end{tabular}
\end{center}
\label{tab:ParamTable}
\end{table}

\begin{figure}[ht!]
\begin{center}
	\includegraphics[width=0.99\columnwidth]{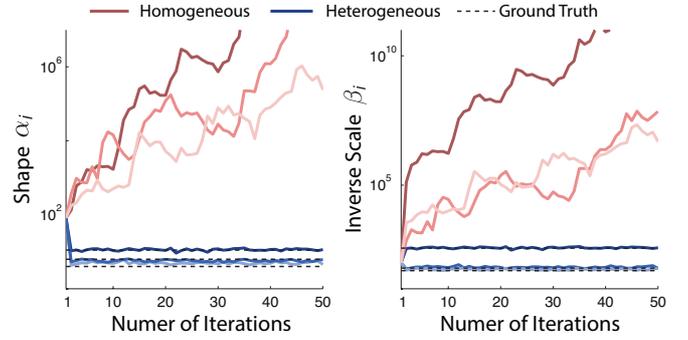}
\caption{Convergence of the VBEM algorithm. The algorithm was applied to $M=30$ cell trajectories between zero and $200min$ with $\kparameter_{3}$, $\kparameter_{5}$ and $\kparameter_{6}$ being heterogeneous with CVs $0.5$, $0.3$ and $0.4$, respectively. The algorithm was ran for $50$ update iterations. The curves correspond to expected values of the respective quantity (i.e., $\alpha_i$, $\beta_i$).}
\label{fig:Figure2}
\end{center}
\end{figure}

We first analyzed convergence of the VBEM algorithm using the network from Fig.~\ref{fig:Model}a and assuming a heterogeneity over three out of the six parameters (i.e., $\kparameters_{3}$, $\kparameters_{5}$ and $\kparameters_{6}$). The results from Fig.~\ref{fig:Figure2} indicate that the algorithm is able to correctly identify the extrinsic noise parameters $\alpha_{i}$ and $\beta_{i}$ in presence of heterogeneity. In case of the homogeneous reactions, both $\alpha_{i}$ and $\beta_{i}$ diverge towards infinity, corresponding to a CV of zero and a finite mean of $\alpha_{i} / \beta_{i}$. Furthermore, we find that in case of the heterogeneous reactions, only very few iterations are necessary until convergence is achieved.

\begin{figure}[ht!]
\begin{center}
	\includegraphics[width=0.99\columnwidth]{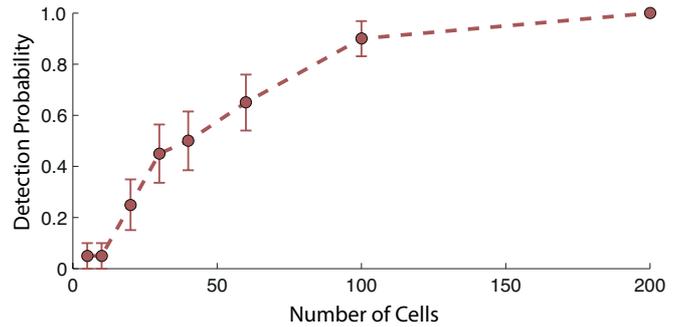}
\caption{Detection robustness as a function of the population size. Probabilities for correct detections were computed for different population sizes (i.e., between $5$ and $200$ cells) using $20$ independent runs. Circles denote mean values and whiskers indicate their standard errors (SEM).}
\label{fig:Figure3}
\end{center}
\end{figure}

Correct identification of the heterogeneous reactions depends on several parameters such as the population size $M$ or the degree of intrinsic noise. In Fig.~\ref{fig:Figure3} we analyze the detection robustness of a single reaction (i.e., the gene-activation event) as a function $M$. In particular, we computed the ratio of positive detections using $20$ independent runs (see figure caption for fuller details). In accordance with our expectations, the results demonstrate that a robust detection of extrinsic variability is possible if enough cells are in place (e.g. around $M>100$ in this case).

\begin{figure}[ht!]
\begin{center}
	\includegraphics[width=0.99\columnwidth]{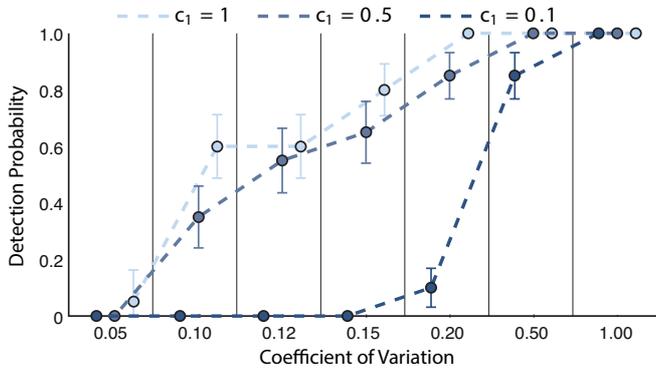}
\caption{Detection robustness as a function of intrinsic and extrinsic noise. We computed the rate of positive detection for different values of $\kparameters_{1}$ yielding different levels of intrinsic noise. For each $\kparameters_{1}$ we computed the detection robustness for several degrees of extrinsic variability (i.e., CVs between $0.05$ and $1$) using $20$ independent runs. Circles denote mean values and whiskers indicate their standard errors (SEM).}
\label{fig:Figure4}
\end{center}
\end{figure}

Similarly, Fig.~\ref{fig:Figure4} shows the probability of successful detection as a function of both \textit{intrinsic} and \textit{extrinsic} variability. Note that intrinsic noise of a reaction firing process scales inversely with its kinetic parameter. Again considering the gene-activation reaction, we computed the detection probabilities for three different values of $\kparameters_1$ (i.e., the intrinsic noise of the expression system) and several degrees of heterogeneity (see figure caption for further details). The parameters $\kparameter_2$ corresponding to the gene-deactivation event was adjusted such as to yield a constant ratio $\kparameter_1 / \kparameter_2$.

We found that in presence of significant intrinsic noise and only moderate degrees of extrinsic noise, the algorithm facilitates the sparsity constraint and hence, yields negative results. In contrast, when decreasing the level of intrinsic noise, the algorithm is widely able to detect the heterogeneity (see Fig.~\ref{fig:Figure4}).

\section{Conclusion}
Recent inference approaches that account for extrinsic variability (\cite{Zechner2012, Zechner2014}) are based on \textit{static} model assumptions, which means that one has to anticipate the events that are heterogeneous among individual cells. In this work we lay out a computational framework to automatically detect the events that are characterized by extrinsic variability using time-lapse data. We show that such a scenario can be understood as a sparse learning problem, which we solve using a variational Bayesian inference scheme. We validate the approach under the simplifying assumption of complete data, generated from a model of eukaryotic gene expression. The framework is currently extended for the use with real-world experimental data.

                                                         
\end{document}